\documentclass[aps,prb,twocolumn,showpacs]{revtex4}
\usepackage[dvips]{graphicx}

\begin{document}

\title{Doping dependence of the electron-doped cuprate superconductors
from the antiferromagnetic properties of the Hubbard model}
\author{Qingshan Yuan,$^1$ Feng Yuan,$^{1,2}$ and C. S. Ting$^1$}
\address{$^1$Texas Center for Superconductivity
and Department of Physics, University of Houston, Houston, TX 77204\\
$^2$Department of Physics, Qingdao University, Qingdao 266071, China}

\begin{abstract}
Within the Kotliar-Ruckenstein slave-boson approach, we have studied
the antiferromagnetic (AF) properties for the $t$-$t'$-$t''$-$U$ model
applied to electron-doped cuprate superconductors.
It is found that, due to the inclusion of quantum fluctuations,
the AF order decays with increasing doping much faster than
obtained in the Hartree-Fock theory. Under an intermediate {\it constant}
$U$ the calculated doping evolution of the spectral intensity in the AF 
state has satisfactorily reproduced the experimental results,
without need of a strongly doping-dependent $U$ as argued 
earlier. This may reconcile a discrepancy in recent studies on
photoemission and optical conductivity.
\end{abstract}

\maketitle
\section{Introduction}
The intriguing Fermi surface (FS) evolution with doping in electron-doped cuprate
Nd$_{2-x}$Ce$_x$CuO$_4$ revealed by angle-resolved photoemission spectroscopy 
(ARPES) measurements\cite{Armitage,Matsui} has attracted much attention
recently.\cite{Kusko,Markiewicz,Kyung,Senechal,Kusunose,Lee,
Yuan04,Yuan05,Tohyama,Luo} It was observed that at low doping a small FS pocket
appears around $(\pi,0)$. Upon increased doping a new pocket begins to emerge
around $(\pi/2,\pi/2)$ and eventually at optimal doping $x=0.15$
the several FS pieces connect to form a large curve around $(\pi,\pi)$.
By use of the $t$-$t'$-$t''$-$U$ model and Hartree-Fock (HF) mean-field 
treatment, Kusko {\it et al.}\cite{Kusko} have studied the FS in consideration
of the antiferromagnetic (AF) order and reached remarkable agreement with
experiments. However, in their work the on-site $U$ is treated as a
doping-dependent effective parameter, specifically $U=6t$ at doping $x=0$
and $3.1t$ at $x=0.15$. Similarly a doping-dependent $U$ was argued 
by others based on numerical calculations.\cite{Kyung,Senechal}
Alternatively, the strong-coupling $t$-$t'$-$t''$-$J$ model was adopted
by some of us to construct the FS.\cite{Yuan04} Without tuning parameters
we have also obtained consistent results with ARPES data.

Very recently, a systematic analysis of the optical conductivity for
electron-doped cuprates was undertaken by Millis {\it et al.}\cite{Millis}
It suggests that (i) the electron-doped materials
are approximately as strongly correlated as the hole-doped ones and 
(ii) the $U$ value is {\it not} strongly doping dependent within the 
electron-doped family. Both of these implications are in apparent disagreement 
with the theoretical studies by Kusko {\it et al.},\cite{Kusko} and in favor of
the strong-coupling model used by us.\cite{Yuan04} On the other hand,
our result at optimal doping from the $t$-$t'$-$t''$-$J$ model\cite{Yuan04}
has not been reconciled by exact diagonalization calculations.\cite{Tohyama}
Thus it is still open whether the strong-coupling model, as implied by
point (i) above, is available to explain the ARPES data at optimal doping.

Here our attention is focused on the issue relevant to point 
(ii) above, i.e., for the $t$-$t'$-$t''$-$U$ model whether the strong doping 
dependence of $U$ as argued by Kusko {\it et al.}\cite{Kusko,Markiewicz} is
really necessary for the interpretation of ARPES data. Theoretically it is not 
convincing to have a strongly doping-dependent $U$ since it is a local 
interaction within an atom and difficult to be changed by adding carriers.
To understand why a strongly doping-dependent $U$ is needed by
Kusko {\it et al.}, we come to some details of their work.
Within the HF mean-field theory, two AF energy bands are derived,
with the minimum of the upper band around $(\pi,0)$ and the maximum of
the lower one around $(\pi/2,\pi/2)$. At low electron-doping,
the upper band is crossed by the Fermi level, leading to a small 
FS pocket around $(\pi,0)$. With increasing doping, the AF order weakens and the 
lower band shifts towards the Fermi level. The essential observation is that
the AF order decays with doping too slowly under a constant $U$,
and consequently the lower band is still too far away from the 
Fermi level even at optimal doping to contribute any spectral intensity
around $(\pi/2,\pi/2)$. Therefore an effective $U(x)$, which decreases with 
increasing $x$, is enforced to expedite the decline of the AF order
so that the lower band approaches the Fermi level
rapidly leading to the consistency with ARPES results.

Actually, as we know, the HF theory overestimates the AF order
due to the ignorance of fluctuations. So a tempting idea is that
the {\it real} AF order should decay with doping in a (much) quicker
way than obtained in the HF theory. Then it is interesting to ask
how much the HF results will be improved if the fluctuations are taken
into account and whether the ARPES data can already be explained with
a {\it doping-independent} $U$.
In this paper we try to answer these questions by studying
the $t$-$t'$-$t''$-$U$ model with the Kotliar-Ruckenstein slave-boson (SB)
approach.\cite{KR} The method well considers the AF fluctuations
for a wide range of interactions\cite{Lilly,Yuan02} and has been
used to study the stripes in hole-doped cuprates.\cite{Seibold} 
We have found that, compared to the HF results, the AF state derived from the
SB approach has a much lower energy and the order parameter decreases much
faster with doping. Consequently, the ARPES results can be reproduced
under a constant $U$.

\section{Formalism}
We start with the $t$-$t'$-$t''$-$U$ model on a square lattice which 
reads
\begin{eqnarray}
H & = & -t\sum_{\langle ij\rangle \sigma}(c_{i\sigma}^{\dagger}c_{j\sigma}+{\rm h.c.})
-t'\sum_{\langle ij\rangle_2\sigma}(c_{i\sigma}^{\dagger}c_{j\sigma}+{\rm h.c.})
\nonumber\\
& & -t''\sum_{\langle ij\rangle_3 \sigma}(c_{i\sigma}^{\dagger}c_{j\sigma}+{\rm h.c.})
+ U\sum_i n_{i\uparrow}n_{i\downarrow}\ ,\label{H}
\end{eqnarray}
where 
$\langle\rangle,\ \langle\rangle_2,\ \langle\rangle_3$
represent the nearest neighbor (n.n.), second n.n., and third n.n. sites, 
respectively, and the rest of the notation is standard.
Throughout the work $t$ is taken as the energy unit and typical parameters
$t'=-0.3$ and $t''=0.2$ are adopted.

In the spirit of the Kotliar-Ruckenstein SB approach,\cite{KR} 
four auxiliary bosons $e_i^{(\dagger)},\ p_{i\sigma}^{(\dagger)}\ 
(\sigma=\uparrow,\downarrow),\ d_i^{(\dagger)}$ are introduced
at each site to label the four different states, which can be
empty, singly occupied by an electron with spin up or down, or
doubly occupied. The advantage of introducing bosons is that the
interaction term can be linearized, i.e., the product of density
operators $n_{i\uparrow}n_{i\downarrow}$ is replaced by
$d_i^{\dagger}d_i$, the occupation number operator for double occupancy. 
As the expense, the hopping term becomes complicated because
any hopping process of electrons must be accompanied by the transitions
of slave bosons. Explicitly, if an electron (with spin $\sigma$)
hops from site $i$ to $j$, the slave bosons must change simultaneously
at both sites $i$ and $j$. For example, at site $i$, associated with
the annihilation of the electron the bosonic state will transit 
either from the singly occupied one with spin $\sigma$
to the empty one or from the doubly occupied one to the singly
occupied one with spin $\bar{\sigma}$. Namely, the transition of the 
bosons at site $i$ can be described by the operator
$z_{i\sigma}=e_i^{\dagger}p_{i\sigma}+ p_{i\bar{\sigma}}^{\dagger}d_i$.
To eliminate the unphysical states in the enlarged (fermionic and bosonic)
Hilbert space, the following constraints have to be imposed at each site:
\begin{eqnarray}
e_i^{\dagger} e_i+\sum_{\sigma} p_{i\sigma}^{\dagger}p_{i\sigma} +
d_i^{\dagger} d_i & = & 1\ ,\label{eq:cons1}\\
p_{i\sigma}^{\dagger} p_{i\sigma} + d_i^{\dagger} d_i & = &
c_{i\sigma}^{\dagger} c_{i\sigma}\ .
\label{eq:cons2}
\end{eqnarray}
The first constraint states that a site is either empty, singly occupied,
or doubly occupied. And the second one guarantees that, when an electron
with spin $\sigma$ locates at site $i$, this site is either singly occupied
with spin $\sigma$ or doubly occupied.
 
In order to study the AF order, we divide the lattice into two sublattices
$A$ and $B$. For sublattice $L$ ($L=A,B$), we introduce a set of bosons 
$e_{iL}^{(\dagger)},\ p_{iL\sigma}^{(\dagger)}\
(\sigma=\uparrow,\downarrow),\ d_{iL}^{(\dagger)}$, and Lagrange multipliers
$\lambda_{iL}^1,\ \lambda_{iL}^{\sigma}$ which are associated
with the constraints (\ref{eq:cons1}) and (\ref{eq:cons2}), respectively.
Thus the original Hamiltonian, with Lagrange multiplier terms added,
can be rewritten, in terms of fermionic operators $a$ and $b$
corresponding to sublattice $A$ and $B$ respectively, as follows
\begin{widetext}
\begin{eqnarray}
H & = & -t\sum_{\langle ij\rangle \sigma,i\in A}
(z_{iA\sigma}^{\dagger} a_{i\sigma}^{\dagger} b_{j\sigma} z_{jB\sigma}+{\rm h.c.})
-t\sum_{\langle ij\rangle \sigma,i\in B}
(z_{iB\sigma}^{\dagger} b_{i\sigma}^{\dagger} a_{j\sigma} z_{jA\sigma}+{\rm h.c.})
-t'\sum_{\langle ij\rangle_2 \sigma,i\in A}
(z_{iA\sigma}^{\dagger} a_{i\sigma}^{\dagger} a_{j\sigma} z_{jA\sigma}+{\rm h.c.})
\nonumber\\
& & -t'\sum_{\langle ij\rangle_2 \sigma,i\in B}
(z_{iB\sigma}^{\dagger} b_{i\sigma}^{\dagger} b_{j\sigma} z_{jB\sigma}+{\rm h.c.})
-t''\sum_{\langle ij\rangle_3 \sigma,i\in A}
(z_{iA\sigma}^{\dagger} a_{i\sigma}^{\dagger} a_{j\sigma} z_{jA\sigma}+{\rm h.c.})
-t''\sum_{\langle ij\rangle_3 \sigma,i\in B}
(z_{iB\sigma}^{\dagger} b_{i\sigma}^{\dagger} b_{j\sigma} z_{jB\sigma}+{\rm h.c.})
\nonumber\\
& & +U\sum_{i\in A} d_{iA}^{\dagger}d_{iA} + U\sum_{i\in B} d_{iB}^{\dagger}d_{iB}
+\sum_{i\in A} \lambda_{iA}^1 (1-e_{iA}^{\dagger} e_{iA}-\sum_{\sigma} 
p_{iA\sigma}^{\dagger}p_{iA\sigma} -d_{iA}^{\dagger} d_{iA})
+\sum_{i\in A,\sigma} \lambda_{iA}^{\sigma}(a_{i\sigma}^{\dagger}a_{i\sigma}
-p_{iA\sigma}^{\dagger}p_{iA\sigma}-d_{iA}^{\dagger} d_{iA})
\nonumber\\ 
& & +\sum_{i\in B} \lambda_{iB}^1 (1-e_{iB}^{\dagger} e_{iB}-\sum_{\sigma}
p_{iB\sigma}^{\dagger}p_{iB\sigma} -d_{iB}^{\dagger} d_{iB})
+\sum_{i\in B,\sigma} \lambda_{iB}^{\sigma}(b_{i\sigma}^{\dagger}b_{i\sigma}
-p_{iB\sigma}^{\dagger}p_{iB\sigma}-d_{iB}^{\dagger} d_{iB})\ ,
\end{eqnarray}
\end{widetext}
where the operators
$$
z_{iL\sigma} =
{e_{iL}^{\dagger}p_{iL\sigma}+p_{iL\bar{\sigma}}^{\dagger}d_{iL}\over 
\sqrt{(1-e_{iL}^{\dagger}e_{iL}-p_{iL\bar{\sigma}}^{\dagger}p_{iL\bar{\sigma}})
(1-d_{iL}^{\dagger}d_{iL}-p_{iL\sigma}^{\dagger}p_{iL\sigma})}}
$$
have been renormalized\cite{KR} to ensure the exact result at $U=0$
even in mean-field treatment as adopted subsequently.
With mean-field approximation,
the bosons are replaced by {\it c} numbers and assumed to be site-independent
on each sublattice, i.e.,
$\langle e_{iL}^{(\dagger)}\rangle=e_{L},\ 
\langle p_{iL\sigma}^{(\dagger)}\rangle=p_{L\sigma},\ 
\langle d_{iL}^{(\dagger)}\rangle=d_{L}$ (and correspondingly 
$\langle z_{iL\sigma}^{(\dagger)} \rangle= z_{L\sigma}$).
At the same time, the constraints 
are softened to be satisfied only on the average on each sublattice,
i.e., $\lambda_{iL}^1 \rightarrow \lambda_{L}^1,\ \lambda_{iL}^{\sigma} 
\rightarrow \lambda_{L}^{\sigma}$.
This treatment is equivalent to making a saddle-point approximation
in the path-integral formulation.
Now the Hamiltonian becomes quadratic in terms of fermionic operators
and can be easily diagonalized in momentum space.
Finally we have
\begin{eqnarray}
H & = & \sum_{k\sigma} (E_{k\sigma}^- \tilde{a}_{k\sigma}^{\dagger}
\tilde{a}_{k\sigma}+E_{k\sigma}^+ \tilde{b}_{k\sigma}^{\dagger} \tilde{b}_{k\sigma})
+E_0\ ,
\end{eqnarray}
where the energy bands read
\begin{widetext}
\begin{equation}
E_{k\sigma}^{\pm} = [\lambda_A^{\sigma}+\lambda_B^{\sigma}+
(z_{A\sigma}^2+z_{B\sigma}^2)\varepsilon'_k]/2 \pm
\sqrt{[\lambda_B^{\sigma}-\lambda_A^{\sigma}+
(z_{B\sigma}^2-z_{A\sigma}^2)\varepsilon'_k]^2/4+
(z_{A\sigma}z_{B\sigma}\varepsilon_k)^2}
\end{equation}
\end{widetext}
with
\begin{eqnarray*}
\varepsilon_k & = & -2t(\cos k_x +\cos k_y)\ ,\\
\varepsilon'_k & = & -4t'\cos k_x\cos k_y -2t'' (\cos 2k_x +\cos 2k_y)\ ,
\end{eqnarray*}
and the constant
\begin{eqnarray*}
E_0 & = & (N/2) \left[ U(d_A^2+d_B^2)
-\lambda_A^1 (e_A^2+\sum_{\sigma}p_{A\sigma}^2+d_A^2-1) \right. \\
& & \ \ \ \ \ \ \ \ \ \ 
-\lambda_B^1 (e_B^2+\sum_{\sigma}p_{B\sigma}^2+d_B^2-1) \\
& & \ \ \ \ \ \ \ \ \ \ \left.
-\sum_{\sigma}\lambda_A^{\sigma}(p_{A\sigma}^2+d_A^2)
-\sum_{\sigma}\lambda_B^{\sigma}(p_{B\sigma}^2+d_B^2) \right]\ .
\end{eqnarray*}
Above $N$ is the total number of lattice sites, $k$ is restricted to the
magnetic Brillouin zone (BZ), and the operators $\tilde{a}$ and 
$\tilde{b}$ are related to $a$ and $b$ by unitary transformations.

The grand canonical thermodynamic potential is given by [$\beta=1/(k_B T)$ with
$T$: temperature]
\begin{equation}
\Omega =  -{1\over \beta}\sum_{k\sigma}\sum_{\nu=+,-}\ln \left[
1+ e^{-\beta (E_{k\sigma}^{\nu}-\mu)}\right]+ E_0\ .
\end{equation}
Here $\mu$ is the chemical potential decided by
$-(\partial \Omega/\partial \mu)=N_e$ ($N_e$ is the total number of electrons).
All the parameters are determined to give the lowest free energy
$F=\Omega + N_e\mu$. The calculation will be simplified if we search for the AF
solution which satisfies the relations: $e_A=e_B,\ d_A=d_B,\ 
p_{A\sigma}=p_{B\bar{\sigma}},\ \lambda_A^1=\lambda_B^1,\ 
\lambda_A^{\sigma}=\lambda_B^{\bar{\sigma}}$.
(Correspondingly the energy bands $E_{k\sigma}^{\pm}$ become spin degenerate,
i.e., $E_{k\sigma}^{\pm}=E_k^{\pm}$.)
The staggered magnetization, which characterizes the AF order,
is given by $m=(p_{A\uparrow}^2-p_{A\downarrow}^2)/2=
(p_{B\downarrow}^2-p_{B\uparrow}^2)/2$.

\section{Results}
In the following we are limited to $T=0$ only at which
the AF long-range order is possible for the two-dimensional model.

\begin{figure}[ht]
\begin{center}
\includegraphics[width=8.5cm,height=4cm,clip]{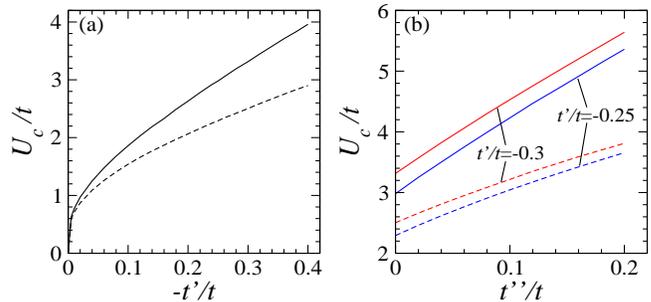}
\end{center}
\caption{(Color online) The critical $U_c$ for the AF order to be stabilized at 
half filling: (a) $U_c$ vs $-t'$ when $t''=0$ (b) $U_c$ vs $t''$ at fixed
$t'/t=-0.25$ [blue (black) lines] and $-0.3$ [red (dark gray) lines]. All solid 
lines are from SB approach and dashed ones from HF theory.}
\label{Fig:Uc}
\end{figure}

To look at the differences between SB and HF results, we first consider
the half-filled case, i.e., $n=N_e/N=1$. As is well known,
the pure $t$-$U$ (with $t'=t''=0$) model takes on the AF order
for arbitrarily small $U>0$ due to the perfect nesting 
effect. After inclusion of the long-range hoppings, the AF order will be 
frustrated and a critical $U=U_c$ is required to stabilize it.
The relations between $U_c$ and $t'$, $t''$
from both SB and HF approaches are compared in Fig.~\ref{Fig:Uc}.
While Fig.~\ref{Fig:Uc}(a) shows $U_c$ vs $t'$ when $t''=0$ which has been given 
previously,\cite{Trapper}
we further present $U_c$ vs $t''$ at fixed $t'$ in Fig.~\ref{Fig:Uc}(b). 
It is clear that $U_c$ increases with both $t'$ and $t''$. Moreover,
the value of $U_c$ at fixed $t'$ and $t''$ obtained from SB approach is always 
larger than that from HF theory. This is expected in view that the HF theory 
overestimates the AF order. Once the fluctuations are taken into account,
the critical value $U_c$ will be largely raised as seen in the SB approach.
We notice that for parameters $t'=-0.25$ and $t''=0.1$ as adopted by
Kusko {\it et al.}\cite{Kusko} the accurate critical value is $U_c=4.23$
from SB approach rather than $3.05$ from HF theory. Thus those values for
doping-dependent $U$ selected by them,\cite{Kusko} i.e.,
$U=3.5 (3.1)$ at $x=0.1 (0.15)$ are actually unreasonable,
which are already too small to stabilize the AF order.

\begin{figure}[ht]
\begin{center}
\includegraphics[width=7cm,height=9cm,clip]{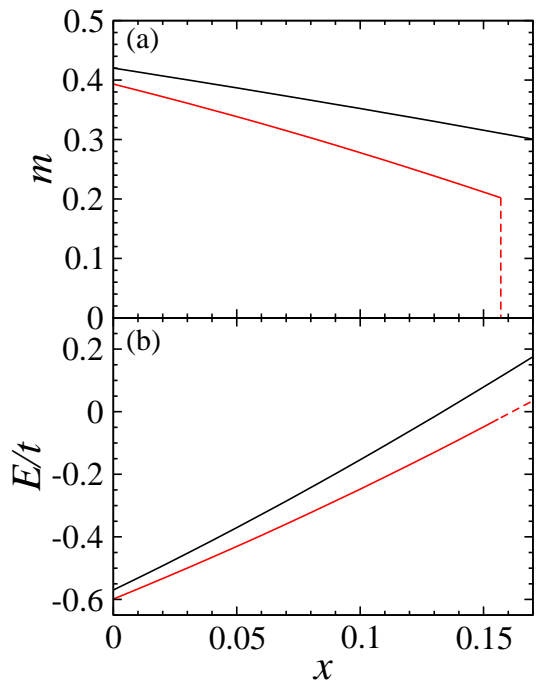}
\end{center}
\caption{(Color online) Staggered magnetization (a) and ground state energy (b) 
vs electron doping for parameters $t'=-0.3$, $t''=0.2$ and $U=6.3$.
The red (dark gray) lines are from SB approach and black ones from HF theory.
The dashed line in panel (a) shows that $m$ jumps to zero at
$x\simeq 0.157$ in SB approach, and correspondingly the dashed one
in panel (b) gives the energy at $m=0$.}
\label{Fig:AF}
\end{figure}

Then we come to see how the AF order declines with increasing doping
$x(=n-1)$ under fixed $U$. For $t'=-0.3$ and $t''=0.2$, the critical value
at half filling is $U_c=5.64$(SB)$/3.81$(HF). We choose $U=6.3>U_c$ and
solve the self-consistent equations. The staggered magnetization $m$ as a function
of $x$ is shown in Fig.~\ref{Fig:AF}(a), where the red (dark gray) and black 
lines are obtained from SB and HF approaches, respectively.
It is seen that at half filling the AF order derived from SB approach becomes
weak relative to that from HF theory. And it is even weaker with increasing
doping because $m$ decreases much faster in SB approach than in HF theory.
In addition, we have found that the AF state
becomes energetically unfavorable at $x\simeq 0.157$ in SB approach,
leading to a first-order transition as shown by the dashed line
in Fig.~\ref{Fig:AF}(a). This first-order termination of the AF order,
compared to the claim of a quantum critical point,\cite{Kyung} seems to be
closer to the experimental observation where a steep disappearance of the
AF phase was found.\cite{Fujita} Also the doping range that the AF order
survives is quantitatively consistent with experiment.\cite{Mang}
In contrast, $m$ keeps finite up to $x\sim 0.4$ and
goes to zero continuously in HF theory (not shown).
The corresponding ground state energies from both approaches are
compared in Fig.~\ref{Fig:AF}(b). The energy derived from SB approach
is notably lower than that from HF theory and the energy difference
increases with increasing doping. This indicates that the study of the AF 
property is largely improved within the SB approach.

\begin{figure}[ht]
\begin{center}
\includegraphics[width=8.5cm,height=6.5cm,clip]{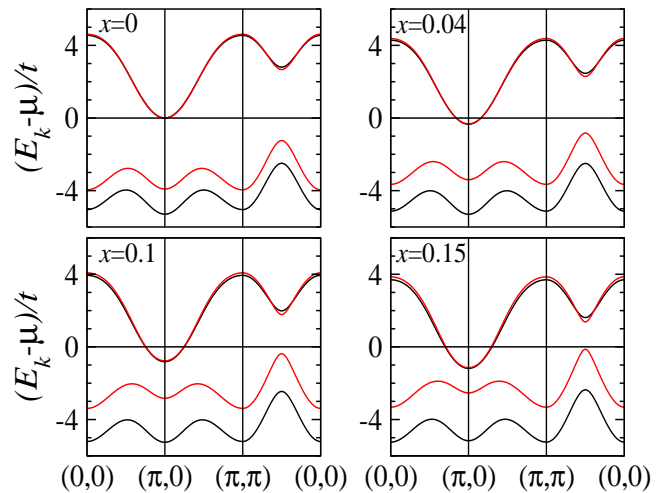}
\end{center}
\caption{(Color online) The energy bands at different dopings. In each panel, 
the red (dark gray) and black lines are from SB and HF approaches, respectively.
The Fermi energy is fixed at zero.}
\label{Fig:band}
\end{figure}

Subsequently the energy bands $E_k^{\pm}-\mu$
are plotted in Fig.~\ref{Fig:band}, with red (dark gray) lines from SB approach
and black ones from HF theory.
(The definitions of $E_k^{\pm}$ in the magnetic BZ have been
analytically extended to the original BZ.)
While the upper bands derived from two approaches look similar at 
each doping, the lower bands take on different behaviors.
Within HF theory, the lower band shifts very slowly towards the Fermi level
with increasing doping and is still far away from it 
at $x=0.15$. On the other hand, within SB approach, the lower band lifts
towards the Fermi level as a whole. At $x=0.1$ it is already
rather close to the latter and may contribute the spectral intensity
within the experimental resolution. At optimal doping, it is nearly crossed
by the Fermi level around $(\pi/2,\pi/2)$.

\begin{figure}[ht]
\begin{center}
\includegraphics[width=8.6cm,height=5.7cm,clip]{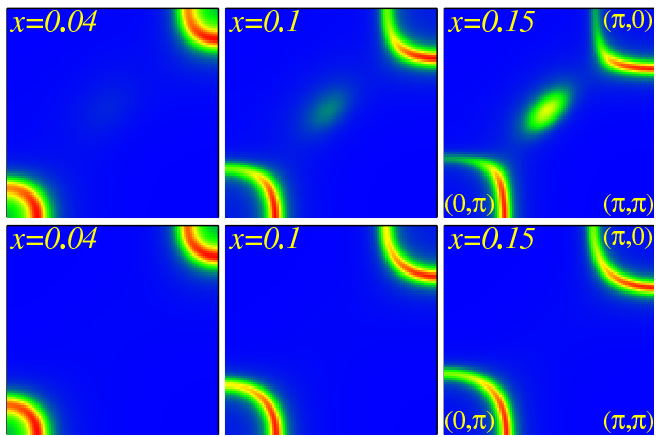}
\end{center}
\caption{(Color online) Density plots for integration of $A(k,\omega)$ times 
the Fermi function over an energy interval $[-0.12,0.06]t \simeq [-40,20]$ meV
around the Fermi level, at doping $x=0.04$, $0.1$ and $0.15$ (from left to right).
The top row is from the SB approach and the bottom row from HF theory,
under a doping-independent constant $U=6.3$ (with $t'=-0.3$ and $t''=0.2$).
The Lorentzian broadening is $0.15t$.}
\label{Fig:Akw}
\end{figure}

In order to directly compare with the ARPES data, we have calculated the
spectral function $A(k,\omega)$ within SB approach.
The density plots for integration of $A(k,\omega)$ times the
Fermi function over an energy interval $[-40,20]$ meV
(same as that adopted in ARPES experiments\cite{Armitage})
around the Fermi level are shown in Fig.~\ref{Fig:Akw} by the top row.
At low doping $x=0.04$ a small FS pocket forms around $(\pi,0)$
[and equivalently $(0,\pi)$]. At doping $x=0.1$ the spectral intensity
begins to appear around $(\pi/2,\pi/2)$, and becomes strong at optimal doping.
At the same time, the Fermi patch around $(\pi,0)$ deforms with increasing doping,
i.e., half of the FS around $(\pi,0)$ loses much of its intensity.
The theoretical results agree well with the ARPES data.\cite{Armitage}
For comparison, the corresponding plots from HF theory are shown by
the bottom row in Fig.~\ref{Fig:Akw}, which, however, do not exhibit the
main experimental features.

\section{Discussions and Conclusion}

In the above sections, after suitable consideration of the AF fluctuations
we have reached consistent results with ARPES measurements under a
doping-independent constant $U$, in contrast with the previous argument for
a doping-dependent $U$.\cite{Kusko,Markiewicz,Kyung,Senechal} The agreement
is satisfactory in view that all input parameters except $U$
are extracted from experiments. We do not exclude that a tunable $U$
with doping will help lead to a perfect comparison between theory
and experiment. Even in this case, however, not a strong doping-dependence
of $U$ as given by Kusko {\it et al.}
($\sim 50$\% variation from $x=0$ to $x=0.15$)\cite{Kusko} is needed.
This may reconcile one of the discrepancies in recent explanations
of both ARPES and optical conductivity mentioned before.

As for the magnitude of $U$, an intermediate value $6.3t$ is adopted here,
which is somewhat too small at $x=0$ to produce the observed Mott gap
($\sim 1$ eV)\cite{Armitage} in the parent compounds of electron-doped 
cuprates. A larger $U\sim 8t$ is expected, but it will be unfavorable to the
appearance of the spectral intensity around $(\pi/2,\pi/2)$ in our current
calculation. On the other hand, we should note that, if the AF fluctuations
are fully taken into account, an even quicker decreasing of $m$ with doping
than displayed in Fig.~\ref{Fig:AF}(a) will be the case. Then a larger $U$
than adopted here, towards strong-coupling regime, could be chosen and
fixed throughout the doping range, leading to the similar intensity maps
shown in the top row of Fig.~\ref{Fig:Akw}.
The evaluation of the biggest $U$ available to explain the ARPES data
as well as the possibility to cross over to the strong-coupling model
need further work.\cite{Dahnken}

In conclusion, we have studied the AF properties for the electron-doped
$t$-$t'$-$t''$-$U$ model by use of the Kotliar-Ruckenstein SB approach.
With inclusion of quantum fluctuations the AF order declines 
with increasing doping much faster than obtained in the HF theory.
The calculated doping evolution of the spectral intensity under
a constant $U$ is in good agreement with ARPES measurements on
electron-doped cuprates. A strongly doping-dependent $U$, 
which is difficult to justify itself, is found to be actually unnecessary.

\section*{ACKNOWLEDGMENTS}
We would thank T. K. Lee and R. S. Markiewicz for useful communications.
This work was supported by the Texas Center for Superconductivity at
the University of Houston and Grant No. E-1146 from the Robert A. Welch
Foundation.


\begin{references}
\bibitem{Armitage} N. P. Armitage, F. Ronning, D. H. Lu, C. Kim, A. Damascelli,
K. M. Shen, D. L. Feng, H. Eisaki, Z.-X. Shen, P. K. Mang, N. Kaneko, M. Greven,
Y. Onose, Y. Taguchi, and Y. Tokura, Phys. Rev. Lett. {\bf 88}, 257001 (2002).
\bibitem{Matsui} H. Matsui, K. Terashima, T. Sato, T. Takahashi, S.-C. Wang,
H.-B. Yang, H. Ding, T. Uefuji, and K. Yamada,
Phys. Rev. Lett. {\bf 94}, 047005 (2005).
\bibitem{Kusko} C. Kusko, R. S. Markiewicz, M. Lindroos, and A. Bansil,
Phys. Rev. B {\bf 66}, 140513(R) (2002).
\bibitem{Markiewicz} R. S. Markiewicz, Phys. Rev. B {\bf 70}, 174518 (2004).
\bibitem{Kyung} B. Kyung, V. Hankevych, A.-M. Dar\'e,
and A.-M. S. Tremblay, Phys. Rev. Lett. {\bf 93}, 147004 (2004).
\bibitem{Senechal} D. S\'en\'echal and A.-M. S. Tremblay,
Phys. Rev. Lett. {\bf 92}, 126401 (2004).
\bibitem{Kusunose} H. Kusunose and T. M. Rice, Phys. Rev. Lett. {\bf 91},
186407 (2003).
\bibitem{Lee} T. K. Lee, C. M. Ho, and N. Nagaosa, Phys. Rev. Lett.
{\bf 90}, 067001 (2003).
\bibitem{Yuan04} Q. S. Yuan, Y. Chen, T. K. Lee, and C. S. Ting,
Phys. Rev. B {\bf 69}, 214523 (2004).
\bibitem{Yuan05} Q. S. Yuan, T. K. Lee, and C. S. Ting,
Phys. Rev. B {\bf 71}, 134522 (2005).
\bibitem{Tohyama} T. Tohyama, Phys. Rev. B {\bf 70}, 174517 (2004).
\bibitem{Luo} H. G. Luo and T. Xiang, Phys. Rev. Lett.
{\bf 94}, 027001 (2005).
\bibitem{Millis} A. J. Millis, A. Zimmers, R. P. S. M. Lobo, and N. Bontemps,
cond-mat/0411172.
\bibitem{KR} G. Kotliar and A. E. Ruckenstein, Phys. Rev. Lett.
{\bf 57}, 1362 (1986).
\bibitem{Lilly} L. Lilly, A. Muramatsu, and W. Hanke, Phys. Rev. Lett. 
{\bf 65}, 1379 (1990).
\bibitem{Yuan02} Q. S. Yuan and T. Kopp, Phys. Rev. B {\bf 65}, 085102 (2002).
\bibitem{Seibold} G. Seibold and J. Lorenzana,
Phys. Rev. B {\bf 69}, 134513 (2004).
\bibitem{Trapper} U. Trapper, H. Fehske, and D. Ihle,
Physica C {\bf 282-287}, 1779 (1997); I. Yang, E. Lange, and G. Kotliar,
Phys. Rev. B {\bf 61}, 2521 (2000). 
\bibitem{Fujita} M. Fujita, T. Kubo, S. Kuroshima, T. Uefuji, K. Kawashima,
K. Yamada, I. Watanabe, and K. Nagamine, Phys. Rev. B {\bf 67}, 014514 (2003). 
\bibitem{Mang} P. K. Mang, O. P. Vajk, A. Arvanitaki, J. W. Lynn, and M. Greven,
Phys. Rev. Lett. {\bf 93}, 027002 (2004).
\bibitem{Dahnken} This possibility seems to be supported from the very recent
numerical study, see C. Dahnken, M. Potthoff, E. Arrigoni, and W. Hanke,
cond-mat/0504618.
\end{references}
\end{document}